\documentstyle[12pt,aaspp4,flushrt]{article}

\newcommand{\kms}{{\rm\ km\ s}$^{-1}$}
\newcommand{\cms}{{\rm\ cm\ s}$^{-1}$}
\newcommand{\pccm}{{\rm\ cm}$^{-3}$}

\newcommand{\etal}{{et al.\,}}

\newcommand{\ha}{H$\alpha$}

\newcommand{\uc}{_{\rm c}}

\def\u#1{_{\rm #1}}

\def\ee#1{\times10^{#1}}

\begin{document}

\title{Jet-cloud interactions and the brightening of  \\ 
the narrow line region in Seyfert galaxies}

\author{
W. Steffen\altaffilmark{1}, J.L. G\'omez\altaffilmark{2},  
A.C. Raga\altaffilmark{3} and R.J.R. Williams\altaffilmark{4}}

\altaffiltext{1}{Department of Physics and Astronomy, University of
Manchester, Manchester M13 9PL, UK; wsteffen@ast.man.ac.uk} 

\altaffiltext{2}{Instituto de Astrof\'{\i}sica de
Andaluc\'{\i}a, CSIC, Apdo. 3004, Granada 18080, Spain;
jlgomez@iaa.es} 

\altaffiltext{3}{Instituto de Astronom\'{\i}a, UNAM, Apdo. Postal
70-264, 04510 M\'exico, D.F., Mexico; raga@astroscu.unam.mx} 

\altaffiltext{4}{Department of Physics and Astronomy, 
         The University, Leeds LS2 9JT, UK; rjrw@ast.leeds.ac.uk
} 

\date{}

\begin{abstract}
  We study the kinematical and brightness evolution of emission line
  clouds in the narrow line region (NLR) of Seyfert galaxies during
  the passage of a jet.  We derive a critical density above which a
  cloud remains radiative after compression by the jet cocoon. The
  critical density depends mainly on the cocoon pressure.
  Super-critical clouds increase in emission line brightness, while
  sub-critical clouds generally are highly overheated reducing their
  luminosity below that of the inter-cloud medium.
  
  Due to the pressure stratification in the bow-shock of the jet, a
  cylindrical structure of nested shells develops around the jet. The
  most compact and brightest compressed clouds surround the cloud-free
  channel of the radio jet. To support our analytical model we present
  a numerical simulation of a supersonic jet propagating into a clumpy
  NLR.  The position-velocity diagram of the simulated \ha\ emission
  shows total line widths of the order of 500\kms\ with large-scale
  variations in the radial velocities of the clouds due to the
  stratified pressure in the bow-shock region of
  the jet. Most of the luminosity is concentrated in a few dense
  clouds surrounding the jet. These morphological and kinematic
  signatures are all found in the well observed NLR of NGC~1068 
  and other Seyfert galaxies.

\end{abstract}


\section{Introduction} 
\label{intro.sec}

Distinct signs of interaction between a collimated radio jet and a
clumpy narrow line region (NLR) are found in some Seyfert galaxies
(e.g.  NGC~1068), such as morphological association between radio and
optical structures (Capetti \etal 1995, Capetti, Macchetto, \&
Lattanzi 1997; Gallimore, Baum,\& O'Dea 1996) and/or in wide spectral
lines of FWHM $\sim$ 1000 \kms\ (Veilleux 1991). Even though
ground-based spatially resolved spectroscopy has yielded some
information on the kinematics of individual clouds and the relation to
the interaction with the radio outflows (Wagner \& Dietrich 1996;
P\'econtal \etal 1997) it is not yet clear whether the high velocities
of the clouds can be attributed to the interaction with the jet. The
new Space Telescope Imaging Spectrograph (STIS) on the Hubble Space
Telescope (HST) should provide the necessary spatial resolution to
test this jet-cloud interaction hypothesis.

Apart from the kinematics we consider the structural changes in the
NLR caused by the interaction with a jet. Linked to this question is
whether the interaction increases or decreases the overall emission
line luminosity, making it more or less likely to observe a jet-NLR
association as compared to an undisturbed NLR. We consider the fate of
clouds with varying density and size after the interaction with the
jet cocoon. This problem is similar to the case of a supernova shock
over-running a cloud in the interstellar medium which has been studied
by several authors (e.g. McKee \& Cowie 1975; Hartquist \& Dyson 1993
and references therein; Klein, McKee, \& Colella 1994). The adiabatic
interaction of a jet with an ensemble of clouds has been studied
numerically by Higgins, O'Brien, \& Dunlop (1995). Bicknell, Dopita,
\& O'Dea (1997) have presented an analytical approach to explain the
relationship between emission-line luminosity and radio-power in
different clases of AGN using a model based on the jet/cocoon
interaction with the ISM. Steffen \etal (1997b) have carried out
non-adiabatic simulations of a jet propagating into a filamentary
medium on the scale of the NLR in Seyfert galaxies.

In this Letter we concentrate on the consequences of the interaction on
a large ensemble of clouds based on analytic and numerical
calculations of the change of the luminosity of the clouds. In
particular, we discuss the effect which the stratification of pressure
in the bowshock has on the kinematics and emission of the ensemble of
clouds.

\section{Critical cloud densities}
\label{analytic.sec}

In this section we derive a minimum number density $n\u{cc}$ for a
cloud above which radiative shock compression occurs after being
entrained or hit by the over-pressured cocoon of the jet. If the
compressing shock is radiative, the luminosity increases, while
otherwise the cloud is overheated and reduces its emission line
brightness. In order to derive the critical density we assess whether
the cooling time $t\u{cool}$ of a compressed cloud of initial density
$n\u{c0}$ is sufficiently small to lead to a dense, cold cloud with a
temperature $T\u{cf}$ similar to its initial value $T\u{c0}$ ($\approx
10^4$K) within a shock crossing time $t\u{cc}$. Hence, we require
$t\u{cool}<t\u{cc}$.

In order to compare the shock crossing time with the cooling time we
assume a cooling function of the form $\Lambda(T)=\Lambda_0
T^{-\alpha}$ (with $\Lambda_0=4.6\ee{-18}{\rm erg~s^{-1}~cm^3}$ and
$\alpha=0.76$ for $T>1.5\ee{5}{\rm K}$; Taylor, Dyson, \& Axon 1992),
where $T$ is the temperature. The cooling of the immediate post-shock
gas can then be expressed in terms of the pressure P driving the
shock, using $T=P/\sigma n\u{c0} k$ (where $\sigma=4$ is the
compression ratio by a strong adiabatic shock and $k$ is the 
Boltzmann constant):\\
\begin{equation}
\label{cool.eq}
\Lambda(P) = \Lambda_0 \left(\frac{P}{\sigma n\u{c0} k}\right)^{-\alpha}.
\end{equation}
The cooling time-scale
$t\u{cool}$ of the post-shock gas is given by the ratio between the
internal energy $\gamma P$ and the cooling rate $(\sigma n\u{c0})^2
\Lambda(P)$, 
\begin{equation}
\label{tau.eq}
t\u{cool} = \frac{\gamma P^{1+\alpha}}
                 {k^\alpha(\sigma n\u{c0})^{2+\alpha} \Lambda_0}
\end{equation}
where we made use of equation (\ref{cool.eq}) and $\gamma = 5/3$ is the
specific heat ratio for a monatomic gas.  The critical density
$n\u{cc}$ of a cloud above which it will cool sufficiently fast to the
photoionization equilibrium temperature is estimated by equating the
cooling time with the time $t\u{cc}=d\u{c0} v\uc^{-1}$ which 
the compressing shock of speed $v\uc$ takes to cross a cloud of size
$d\u{c0}$.  The propagation speed of the shock in a cloud of
number density $n\u{c0}$ can be calculated using
\begin{equation}
\label{shvel.eq}
v\uc = \left(\frac{\xi P}{\mu n\u{c0}}\right)^\frac{1}{2},
\end{equation}
where $\mu$ is the average mass per atom or ion and $\xi=1$ for a
strong isothermal shock and $\xi=4/3$ otherwise.

Combining equations (\ref{tau.eq}) and (\ref{shvel.eq}) with the shock
crossing time we find the critical number density $n\u{cc}$ of the
unshocked cloud as a function of the compressing pressure and
initial cloud size
\begin{equation}
\label{critn.eq}
n\u{cc} = \left( \frac{\xi^{1/2} \gamma P^{3/2+\alpha}} 
              {\sigma^{2+\alpha} k^\alpha \Lambda_0 \mu^{1/2} d\u{c0}}
              \right)^{\frac{1}{\alpha+5/2}}.
\end{equation}
The fact that there is a critical density for the brightening of a
cloud, which depends mainly on the pressure in the cocoon, has important
consequences which could be used to test the significance of the jet
for the observed structure of the NLR. Note that the cocoon pressure
(including ram-pressure) near the head of the bow-shock of the jet
gradually decreases from the axis outwards and from the head towards
the source of the jet, until it reaches a roughly constant
value. Therefore the critical density for the clouds to survive the
passage of the cocoon shock decreases with distance from the jet axis.
Consequently, low-density clouds and filaments will not survive near
the jet axis, resulting in an cylindrical stratification of clouds of
different compression around the jet. This structure is illustrated in
the schematic diagram of Figure \ref{onion.fig} which shows the
inferred stratification as seen along the jet axis. If a cloud is
inhomogeneous, the sub-critical envelope will be stripped off leaving
only any super-critical core. The compactness of the clouds
increases towards the axis, but no clouds will be found at least over
a full jet radius.

\section{Luminosity change of shocked NLR clouds}
\label{lum.sec} 

We now estimate the brightening of a super-critical cloud ($n\u{c0} >
n\u{cc}$) with uniform densities $n\u{c0}$ and $n\u{cf}$ before and
after the compression, respectively. The luminosity ratio is then
\begin{equation}
\label{lum1.eq}
\frac{L\u{cf}}{L\u{c0}} = \frac{\epsilon\u{cf} V\u{cf} }{\epsilon\u{c0} V\u{c0}
}
\end{equation}
where $\epsilon \propto n^2$ and $V$ are the volume emissivity and the
volume of the cloud, respectively. Using the assumption that no mass $M$
($V=M/n$) is lost during the compression (e.g. by ablation or
fragmentation), we obtain from the previous equation
\begin{equation}
\label{lum2.eq}
\frac{L\u{cf}}{L\u{c0}} = \frac{n\u{cf}}{n\u{c0}}.
\end{equation}
The final compression ratio $n\u{cf}/n\u{c0}$ can be determined
independently considering that the final pressure of the cloud is the
same as the cocoon pressure $P$ driving the shock into the cloud.  
Using $P = n\u{cf} k T\u{cf}$ and $T\u{cf}\approx T\u{c0} $ together
with equation (\ref{lum2.eq}) we find a very simple expression for the
final luminosity ratio in terms of the pressure of the cloud before
and after the compression
\begin{equation}
\label{clumratio.eq}
\frac{L\u{cf}}{L\u{c0}} = \frac{P}{n\u{c0} k T\u{cf}}
= \frac{P}{P\u{c0}}
\end{equation}
where $P\u{c0}$ is the initial pressure of the cloud. The luminosity
increase of the shocked gas in a super-critical cloud as compared to
its undisturbed value varies as the change in pressure which it
experiences. In the presence of magnetic fields, the compression might
be limited by the magnetic pressure of the compressed cloud rather
than the thermal pressure at the photoionization equilibrium
temperature. If the cloud is too large to be fully processed in the
time-scale of the jet propagation, the increase in brightness will of
course apply only to the shocked fraction of the cloud. If the final density
is above the critical density for collisional deexcitation, the 
emissivity is no longer proportional to $n^2$ and the increase in 
luminosity is different to the one predicted by our simple model, which
nevertheless predicts the correct qualitative result. 

A similar calculation for an adiabatic shock 
yields the luminosity ratio for a sub-critical cloud before and
after compression
\begin{equation}
\label{ulumratio.eq}
\frac{L\u{uf}}{L\u{u0}} = \sigma^{\frac{5}{2}}
                          \left( \frac{P}{P\u{c0}} \right)^{-\frac{3}{2}}
\end{equation}
Here we used equations~(\ref{lum1.eq}) and (\ref{lum2.eq}) with compression
ratio $\sigma$ and volume emissivity $\epsilon\propto n^2 T^{-3/2}$.
The scaling of the emissivity with temperature follows the dependence
of the recombination coefficient. The initial and final temperatures
are assumed to be $T\u{c0}=P_0 / k n\u{c0}$ and $T\u{cf}=P/k\sigma
n\u{c0}$.  Equation~(\ref{ulumratio.eq}) shows that the sub-critical
clouds generally lose flux, since the pressure ratio can be expected
to be much higher than the compression ratio.  Moreover, one finds
that their specific emissivity falls below that of the environment
$\epsilon\u{x}$, in which they where embedded, by a factor of
\begin{equation}
\label{ueps.eq}
\frac{\epsilon\u{uf}}{\epsilon\u{x}} = \sigma^{\frac{7}{2}}
               \left( \frac{P}{P\u{c0}} \right)^{-\frac{3}{2}}
\end{equation}
This basically renders the sub-critical clouds invisible in emission
lines which are emitted most strongly near the equilibrium
temperature.  Even though the cooling time of the hot clouds is much
larger than the internal shock crossing time, it may be of order of
the jet propagation time through the NLR ($10^{12}-10^{14}$~sec).
Consequently these clouds may start radiating after a time of this
order of magnitude and produce emission lines radiated at temperatures
higher than the equilibrium temperature near $10^4$~K with relatively
large velocities. Even though in our numerical simulation (see below)
the emission measure of the hot gas ($3\times 10^4$ to $3\times 10^5$
K) is found to be very low compared to that at lower temperatures
(5000 to 30000 K) (ratio $= 1/530$), this emission, together
with the observed radio emission, could serve as an indicator of the
propagation time of the jet, displaying a variation of temperature and
hence the excitation of the clouds as a function of distance from the
central AGN.

If the NLR is made purely of sub-critical gas, the jet will plough a
cylindrical hole into it. However, if there is a sufficiently large
number of super-critical clouds present, a region of increased
emission line brightness will trace the cocoon of the jet.  If the
average density of the NLR clouds falls with distance from the centre,
there will be a rather sharp transition region beyond which the clouds
are all sub-critical. The consequence will be that the number of
strong shocked NLR clouds will sharply decrease in this region, even if
the jet has already propagated further out, as is observed, e.g. in NGC4151
(Boksenberg \etal 1995; Pedlar \etal 1993) and NGC~1068 (Gallimore
\etal 1996). 

For the NLR to increase in total luminosity due to the
passage of the jet, the ratio between the mass $M_{c0}$ concentrated
in super-critical clouds of uniform density $n\u{c0}$ and the mass
$M_{u0}$ of sub-critical clouds of uniform density $n\u{u0}$ does not
need to be very large, but approximately only
\begin{equation}
\label{mratio.eq}
\frac{M_{c0}}{M_{u0}} > \frac{P\u{c0}}{P}
                  \left(\frac{n\u{u0}}{n\u{c0}}\right)^{9/2}. 
\end{equation}
Equation (\ref{mratio.eq}) shows that the change in overall luminosity
of the NLR depends strongly on the initial average density and mass of
the affected clouds. Only if the total mass of sub-critical clouds is
much larger than the mass of the super-critical ones, can the total
luminosity be comparable for both ensembles. They could contribute 
noticably to the low brightness, but fast wings observed in the emission
lines of many Seyferts. Most of the luminosity though is likely to
be concentrated in a few dense clouds relatively near the jet axis following
the galactic rotation with disturbances of the order of 100-300~\kms.
This is largely consistent with observations of NGC~1068 (Wagner \& Dietrich
1996) and NGC~4151 (Winge \etal 1997).

If initially neutral material is overrun by the cocoon, the increase
in volume emissivity of the shocked material can be expected to be
much larger than the one predicted to occur in preionized clouds,
since the neutral clouds do not emit appreciably in forbidden lines.
They might well be at least partially pre-ionized by the radiation
coming from the cocoon shock (e.g. Bicknell, Dopita, \& O'Dea 1997),
leading to two qualitatively different stages of increase in brightness,
the first being due to precursor photo-ionization and the second due to
ionization and compression by the cocoon shock. 
Consequently, in a statistical sample of Seyfert galaxies,
positional associations between radio jets and emission line gas could
be biased towards galaxies with more neutral gas in the region of
interaction.

Other expected observable consequences of the stratified structure
include a radial variation of the cloud kinematics (faster near the
jet axis) and of the ionization properties of the clouds (since the
column density of neutral ions in the clouds increases due to the
higher recombination rate in the compressed clouds). Possibly, even
polarization properties of scattered nuclear radiation through the
compression of seed magnetic fields might be observable, since the
ram-pressure in the bow-shock will produce enhanced compression
perpendicular to the surface of the bow-shock.

\section{Numerical simulation}
\label{simulation.sec}

In order to test our analytical model we carried out two-dimensional
numerical hydrodynamic simulations in slab-geometry with the code
described in Raga \etal (1995) and Steffen \etal (1997a). The code
uses the flux-vector slitting scheme by van~Leer (1982) on a 5-level
binary adaptive grid. Initially the NLR is stationary and is assumed
fully ionised throughout.  The intercloud medium (ICM) of the NLR has
a number density of 1\pccm and a pressure of $2\times10^{-12}$
dyn~cm$^{-2}$.  The spherical NLR clouds have random positions, radii
in the range $0.5-1.5\ee{19}{\rm cm}$ and a number density of up to
500\pccm (cloud pressure of $1.4\times10^{-9}$ dyne~cm$^{-2}$ for
$n=500$). Most of the mass is concentrated in the clouds (mass ratio
between clouds and ICM is 8.7). The jet has a velocity of $6\ee{9}$\cms,
a number density of 0.3\pccm, and a radius of $8\ee{18}{\rm cm}$.  The
cooling is treated as described in Steffen \etal (1997a), with a lower
limit set to $10^4$K to simulate photoionization equilibrium
(ionization presumably from a the nuclear radiation source).  The grid
size is $513^2$ cells or $(10^{21}{\rm cm})^2$.

The results from the simulation are shown in Figure \ref{sim.fig},
where greyscale images show the distributions of density (A), the \ha\ 
emissivity (B), the initial density (C) and the position-velocity
diagrams (D) of the \ha\ emissivity as seen from directions along
($0^\circ$) and perpendicular ($90^\circ$) to the axis of the jet.
Figure \ref{spec.fig} shows the integrated \ha\ spectrum. Panel (B)
also shows a contour of high temperature ($3\times10^9$ K) indicating
the region where the jet plasma is located and hence from where any
radio emission can be expected to originate, even though the
temperature cannot be a direct measure for the radio intensity. For 
a short time the jet was bent towards the top-right corner, while
in the image shown, it bends slightly downwards. 

The jet propagates into the inhomogeneous NLR shocking the clouds and
the ICM over a large area with a roughly uniform cocoon pressure. The
clouds are entrained into the cocoon and compressed in the way
discussed in Section \ref{analytic.sec}. Note that in a fully
three-dimensional simulation, the transvers extent of the pressurized
region would be somewhat smaller than in this slab-symmetric
simulation. The clouds marked with open arrows in Panel C are
sub-critical clouds and are destroyed during compression by the cocoon
pressure (see same positions marked in Panel A).  The other clouds are
super-critical and turn into discrete, strongly compressed density
concentrations at or near the original position. There are, however, a
number of exceptions, which we have marked with filled arrows. These
clouds are super-critical to the cocoon pressure, but were located
either directly in the path of the jet or close to the jet axis, where
the pressure (and therefore the critical density) was higher when the
interaction started, leading to the destruction of the clouds.

The compactness and brightness of the surviving clouds increases
towards the path of the jet. The increased brightness near the jet can
be seen in Panel B, where the \ha\ emission is shown convolved with a
gaussian of FWHM of 10~pc.  Here we marked three zones. First, a zone
of fully processed compact clouds of highest brightness (a). This
region is surrounded by zone (b) where clouds are less bright and
compact or have been only partially processed (see Panel A), while
clouds in the outer zone (c) have only started the interaction or are
still untouched by the jet cocoon showing very low brightness. This
cylindrical shell structure resembles very much the one observed in
NGC~1068 (Capetti \etal 1997) and NGC~4151 (Winge \etal 1997). The
objects show zones of avoidance of NLR clouds, coincident with radio
features, surrounded by high brightness clouds and weaker filamentary
emission further out.

Compared to the initial conditions, we find that the peak emissivity
rose by approximately three orders of magnitude, strongly emphasizing
a few very dense clouds which emit most of the emission (see the total
spectrum in Figure \ref{spec.fig}). The kinematical disturbance caused
by the passage of the jet through the NLR is most pronounced near the
head of the jet. Most of the strongly emitting gas moves at velocities
under 300\kms, producing a linewidth of around 600\kms, enhanced by
galactic rotation or initial turbulence, which have not been taken
into account in this simulation. Some cloud material with low
emissivity however has been accelerated to speeds of 500\kms and
higher. Seen along a line of sight perpendicular to the jet (Panel D,
$90^\circ$), the pv-diagram appears symmetric. Here the amount of gas
approaching and receding from the observer is well balanced, since the
clouds are compressed equally from all sides without much acceleration
perpendicular to the jet. Viewed along the axis of the jet (Panel D,
$0^\circ$) the acceleration of the clouds due to the interaction with
the bowshock shows as a positive velocity centroid of the cloud
emission (see also Figure \ref{spec.fig}).  There is a gradient of
increasing velocity of the clouds towards the axis of the jet,
consistent with expectations from the stratified accelerating
(ram-)pressure in the bow-shock region.

The morphological and kinematical characteristics of our model are
very similar to those observed in NGC~1068 (e.g. Capetti \etal 1997;
Wagner \& Dietrich 1996) and other NLR of Seyfert galaxies with
associated radio jets and we suggest
that the entrainment and compression of NLR clouds by the 
stratified cocoon pressure of a supersonic jet is the mechanism
dominating the brightness and kinematics of the NLR in this galaxy.
Spectral observations with high spatial resolution, like those of
NGC~1068 presented by P\'econtal \etal (1997) or of NGC~4151 using
HST by Winge \etal (1997), will soon provide sufficient data
to test the general applicability of our model, searching for the
predicted stratified kinematic and structural signatures of the
jet-NLR interaction.

\section{Acknowledgements}
WS acknowledges a PPARC research associateship and is grateful for
the hospitality received by members of the Instituto de
Astrof\'{\i}sica de Andaluc\'{\i}a during which part of this research
was done. RJRW acknowledges a PPARC advanced fellowship. 
This research was partially supported by Spanish DGICYT
(PB94-1275).

\newpage

\begin{figure}
\caption{A schematic view of the radial stratification of clouds after
the entrainment by the cocoon of a supersonic extragalactic jet as seen
along the jet axis. The compactness of the surviving clouds decreases with
increasing distance from the jet axis.}
\label{onion.fig}
\end{figure}

\begin{figure}
\caption{Greyscale distributions are shown for the final
  density (A, logscale), the \ha\ emissivity and temperature contours 
  (B, linear scale), the
  initial number density (C, logscale) and the \ha\ 
  position-velocity diagrams (D, rooted greyscale and contours) as
  seen from directions along ($0^\circ$) and perpendicular
  ($90^\circ$) to the axis of the jet. In the top pv-diagram positions
  along the jet axis are aligned with those in Panel B. For easier
  comparison with observations, the \ha-emissivity map and the
  pv-diagrams have been convolved with a Gaussian
  point-spread-function of FWHM of 10~pc and 60\kms\ in space and
  velocity, respectively.}
\label{sim.fig}
\end{figure}

\begin{figure}
\caption{The integrated \ha\ line profiles for viewing angles which are 
aligned ($0^\circ$) and perpendicular ($0^\circ$) to the jet axis. The peak
at zero-velocity is mainly due to undisturbed gas.}
\label{spec.fig}
\end{figure}

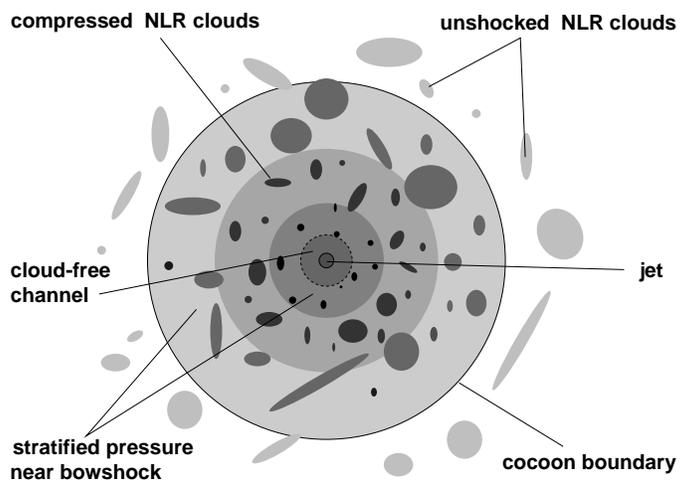
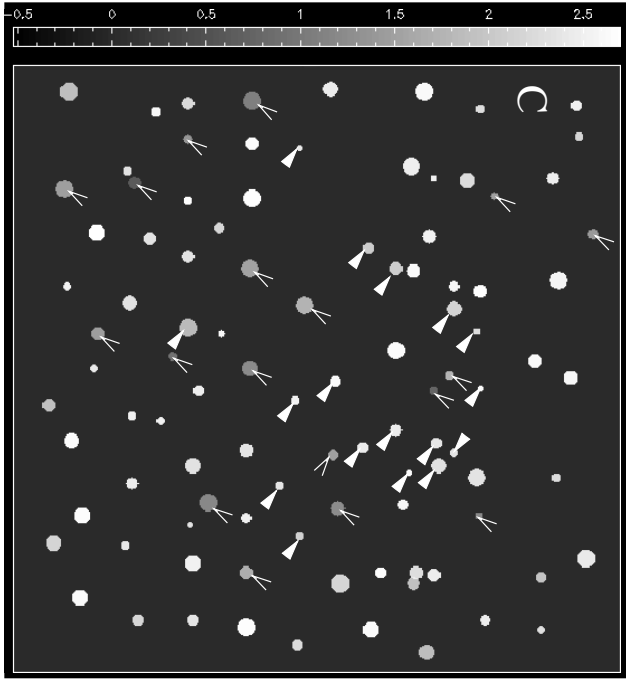
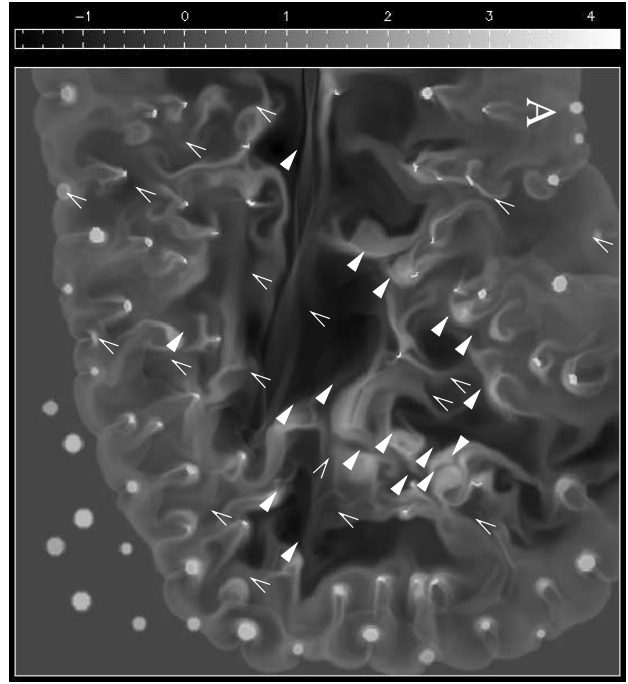
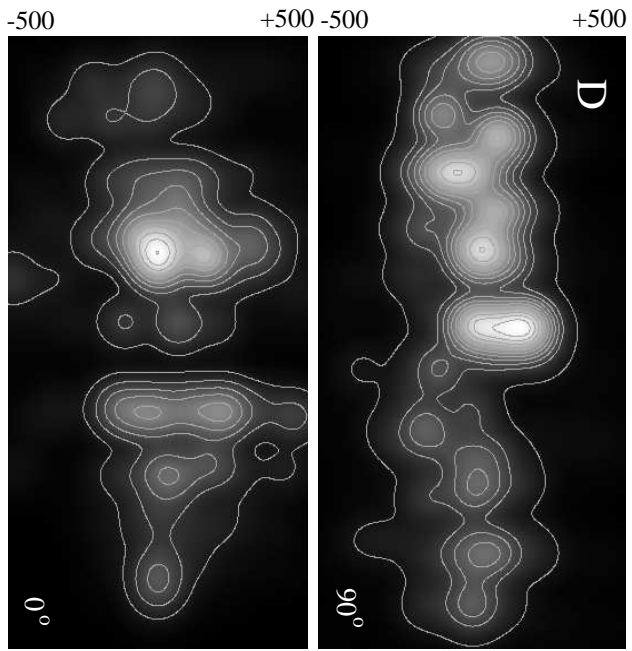
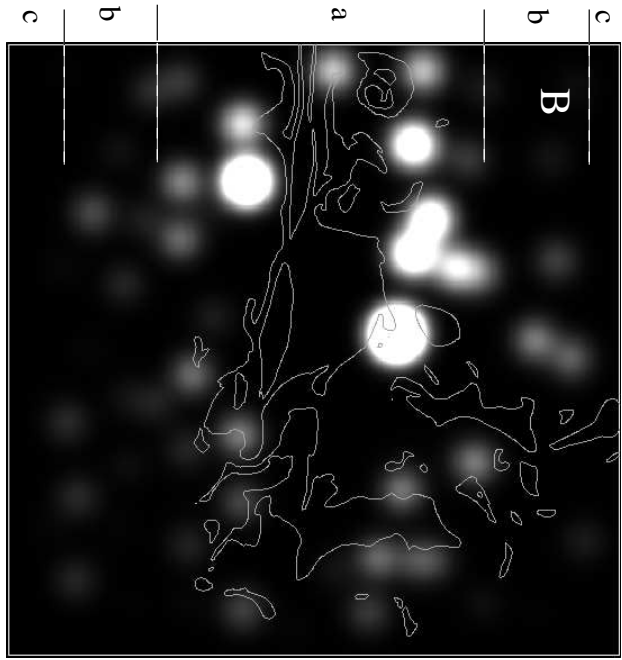
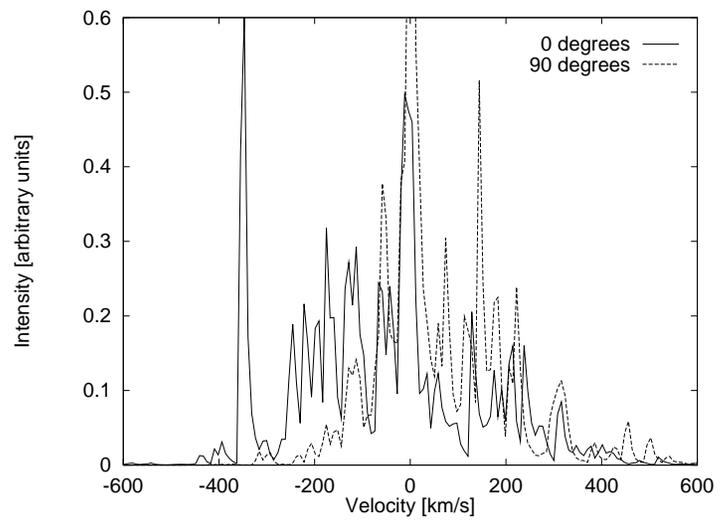
\end{document}